\documentclass{PoS}

\title{Gamma rays from microquasars Cygnus X-1 and Cygnus X-3}

\ShortTitle{Gamma rays from microquasars Cygnus X-1 and Cygnus X-3}

\author{\speaker{A.~Fern\'andez-Barral}\\
        Institut de Fisica d'Altes Energies (IFAE), The Barcelona Institute of Science and Technology (BIST)\\ 
        Campus UAB, 08193 Bellaterra (Barcelona), Spain\\
        E-mail: \email{afernandez@ifae.es}}

\author{O.~Blanch$^{a}$, E.~de O\~na Wilhelmi$^{b}$, D.~Galindo$^{c}$, J.~Herrera$^{d}$, M.~Rib\'o$^{c}$, J.~Rico$^{a}$, A.~Stamerra$^{e}$ (for the MAGIC Collaboration), F.~Aharonian$^{f,g}$, V.~Bosch-Ramon$^{c}$ and R.~Zanin$^{f}$\\
		$^{a}$ IFAE-BIST, Campus UAB, 08193 Bellaterra (Barcelona), Spain\\
		$^{b}$ CSIC/IEEC, E-08193 Barcelona, Spain\\
		$^{c}$ Departament de F\'sica Qu\`antica i Astrof\'isica, Institut de Ci\`encies del Cosmos (ICCUB), Universitat de Barcelona, IEEC-UB, Barcelona, Spain\\
		$^{d}$ IAC and Universidad de La Laguna, E-38200/E-28206 La Laguna, Tenerife, Spain\\
		$^{e}$  INAF - National Institute for Astrophysics, I-00136 Rome, Italy\\
		$^{f}$ Max-Planck-Institut fur Kernphysik, 69029 Heidelberg, Germany\\
		$^{g}$ Dublin Institute for Advanced Studies, Dublin 2, Ireland \\}
        
\abstract{Gamma-ray observations of microquasars at high and very-high energies can provide valuable information of the acceleration processes inside the jets, the jet-environment interaction and the disk-jet coupling. Two high-mass microquasars have been deeply studied to shed light on these aspects: Cygnus X-1 and Cygnus X-3. Both systems display the canonical hard and soft X-ray spectral states of black hole transients, where the radiation is dominated by non-thermal emission from the corona and jets and by thermal emission from the disk, respectively. Here, we report on the detection of Cygnus X-1 above 60 MeV using 7.5 yr of \texttt{Pass8} Fermi-LAT data, correlated with the hard X-ray state. A hint of orbital flux modulation was also found, as the source is only detected in phases around the compact object superior conjunction. We conclude that the high-energy gamma-ray emission from Cygnus X-1 is most likely associated with jets and its detection allow us to constrain the production site. Moreover, we include in the discussion the final results of a MAGIC long-term campaign on Cygnus X-1 that reaches $\sim100$ hr of observations at different X-ray states. On the other hand, during summer 2016, Cygnus X-3 underwent a flaring activity period in radio and high-energy gamma rays, similar to the one that led to its detection in the high-energy regime in 2009. MAGIC performed comprehensive follow-up observations for a total of $\sim70$ hr. We discuss our results in a multi-wavelength context.}

\FullConference{35th International Cosmic Ray Conference  -ICRC217-\\
		10-20 July, 2017\\
		Bexco, Busan, Korea}

\begin{document}

\section{Introduction}
Cygnus\,X-1 is an X-ray binary comprised by a (19.2$\pm$1.9) M$_{\odot}$ O9.7Iab supergiant star and a (14.8$\pm$1.0) M$_{\odot}$ BH \cite{Orosz2011}, classified as a microquasar after the detection of a one-sided relativistic radio-jet \cite{Stirling2001}. The jet seems to create a 5 pc ring-like structure detected in the radio that extends up to $10^{19}$ cm from the BH \cite{Gallo2005}. The system follows an almost circular orbit of $\sim 5.6$ d period \cite{Brocksopp1999a}. Flux modulation with the orbital period is detected in X-ray and radio \cite{Wen1999,Brocksopp1999b, Szostek2007}, produced by the absorption/scattering of the radiation by the stellar wind. 
Cygnus\,X-1 displays the two principal X-ray states of BH transients, the soft state (SS) and the hard state (HS). Both are described by the sum of a blackbody-like emission from the accretion disk that peaks at $\sim 1$ keV (dominant in the SS) and a power-law tail with a cutoff at hundred keV, expected to be originated by inverse Compton (IC) scattering on disk photons by thermal electrons in the so-called \textit{corona} (dominant in the HS). During HS the source displays persistent jets from which synchrotron radio emission is detected, whilst in the SS, these jets are disrupted. Cygnus\,X-1 showed a $4\sigma$-hint above 100 MeV during HS reported by \cite{Malyshev}, using 3.8 yr of \textit{Fermi}-LAT data. Evidences of flaring activity were also reported by \textit{AGILE} ($> 100$ MeV, \cite{AGILE2010Sabatini, AGILE2010Bulgarelli,AGILE2013}) and by MAGIC ($> 100$ GeV, \cite{Albert2007}). 

The microquasar Cygnus\,X-3 hosts a Wolf-Rayet (WR) star, although it follows a short 4.8 hr-orbit. The compactness of the system produces an unusually high absorption, which complicates the identification of the compact object (1.4 M$_{\odot}$ neutron star (NS) \cite{Stark2003} or $< 10$ M$_{\odot}$ BH \cite{Hanson2000}). Despite this high absorption, its X-ray spectrum shows the two aforementioned states. Cygnus\,X-3 is the strongest radio source among the X-ray binaries, whose flux can vary several orders of magnitude during its frequent radio outbursts. These major flares happen only during SS (see \cite{Szostek2008}). Cygnus\,X-3 was detected above 100 MeV, during SS by AGILE \cite{Tavani2009} and \textit{Fermi}-LAT \cite{Fermi2009}. Its spectrum was described as a power law with photon indices 1.8$\pm$0.2 and $2.70\pm0.25$, respectively. 

Here, we present the results for GeV and TeV searches on Cygnus\,X-1 using 7.5yr of \textit{Fermi}-LAT data and $\sim 100$ hr of MAGIC data. We also show the latest results of Cygnus\,X-3 obtained with MAGIC during the August-September 2016 flare. 

\section{Observations and Analysis}
\textit{Fermi}-LAT is the principal scientific instrument on the Fermi Gamma-ray Space Telescope spacecraft that studies the gamma-ray sky within an energy range of $\sim 20$ MeV to $\sim 500$ GeV (see \cite{PerformanceFermi}). To study Cygnus\,X-1 in the high-energy (HE; $>60$ MeV) regime, we used 7.5 years of \texttt{Pass8} \textit{Fermi}-LAT data (from MJD 54682--57420). The analysis was performed using \textit{Fermipy}\footnote{http://fermipy.readthedocs.io/en/latest/}, a package of python tools to automatize the analysis with the FERMI SCIENCE TOOLS (v10r0p5 package). We selected photon-like events between 60 MeV and 500 GeV, within a 30$^{\circ}$ radius centered at the position of Cygnus\,X-1. Find more details in \cite{Zanin2016}.

MAGIC is a stereoscopic system of two 17 m diameter Cherenkov Telescopes located in La Palma (Spain). Until 2009, MAGIC consisted in just one telescope \cite{Aliu2009}. After autumn 2009, MAGIC\,II started operation \cite{Alecksic2012} and between 2011-2012, both telescopes underwent a major upgrade \cite{Alecksic2016}. MAGIC observed Cygnus\,X-1 for $\sim 100$ hours between 2007 and 2014 mostly during its HS (see \cite{FernandezBarral2017}). This analysis was carried out with standard MAGIC software (MARS, \cite{Zanin2013}). Upper limits (ULs) at 95\% confidence level (CL) were computed with the full likelihood analysis developed by \cite{AleksicLikelihood}, assuming 30\% systematic uncertainty. 

Between August and September 2016, Cygnus\,X-3 experienced strong flaring activity in radio and HE regimes during its SS \cite{RadioATel, FermiATel}. MAGIC observed the source $\sim 70$ hours between MJD 57623 to 57653, under different moonlight conditions (moon analysis performed following \cite{MoonPerformance}). ULs at 95\% CL were computed following Rolke method \cite{Rolke2005}. 

\section{Results}
\subsection{Cygnus\,X-1}
\begin{center}
\begin{figure}
\centering
		\includegraphics[width=0.9\linewidth]{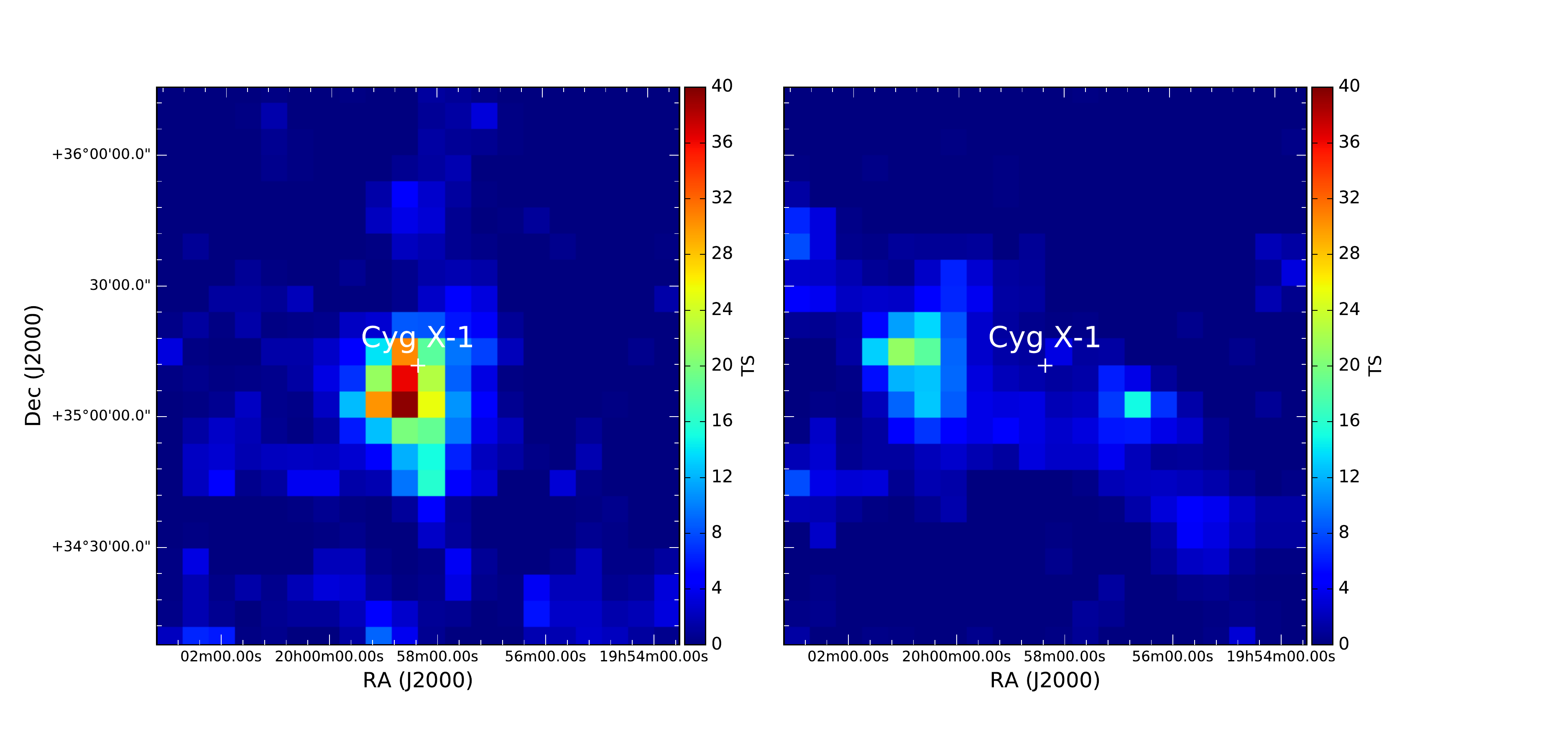}
		\caption{TS maps above 1 GeV centered in Cygnus\,X-1, using HS (\textit{left}) and SS subsamples (\textit{right}).}
		\label{FermiSkypmaps}
\end{figure}
\end{center}

\begin{center}
\begin{figure}
\centering
		\includegraphics[width=0.7\linewidth]{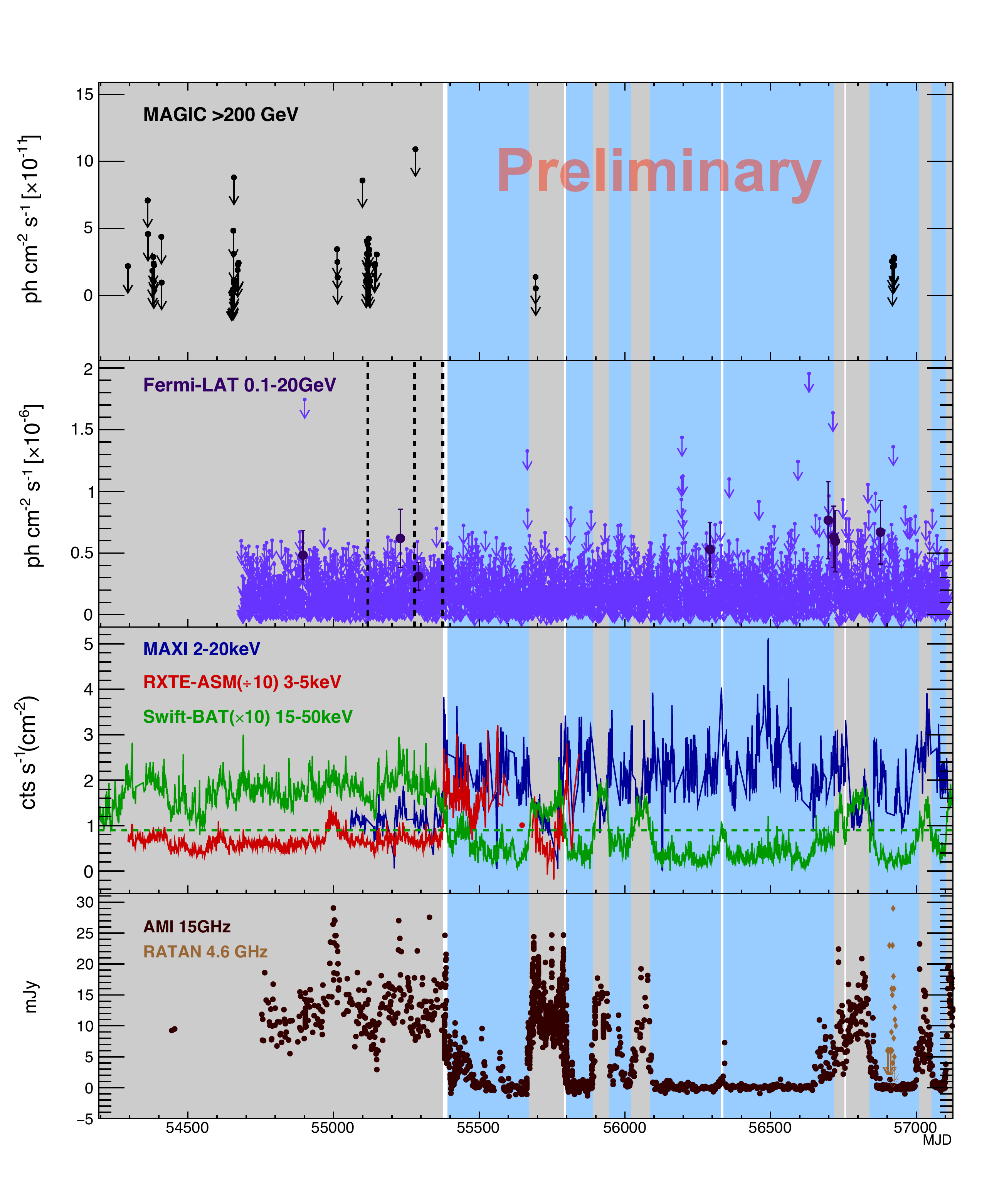}
		\caption{Multi-wavelength light curve for Cygnus\,X-1. \textit{From top to bottom:} Daily MAGIC ULs ($> 200$ GeV), HE gamma rays from the \textit{Fermi}-LAT analysis (flux points are computed when $TS>9$), hard X-rays from \textit{Swift}-BAT (15-50 keV, \protect\cite{Krimm2013}), soft X-rays from MAXI (2--20 keV, \protect\cite{Matsuoka2009}) and \textit{RXTE}-ASM (3--5 keV range), and radio from AMI (15 GHz) and RATAN-600 (4.6 GHz). In the HE pad, dashed lines correspond to \textit{AGILE} transient events. The horizontal green line in \textit{Swift}-BAT pad shows the limit at 0.09 cts cm$^{-2}$ s$^{-1}$ given by \protect\cite{Gringberg2013} to differentiate between X-ray states. HS and SS periods are highlighted with grey and blue bands, respectively.}
		\label{CygX1LC}
\end{figure}
\end{center}

\textit{Fermi}-LAT skymap, between 60 MeV and 500 GeV, showed a point-like source at the position of Cygnus\,X-1 with a TS=53. Moreover, detection only happens during HS (Figure \ref{FermiSkypmaps}) with TS=49 above 60 MeV (division between HS and SS done following \cite{Gringberg2013}). Therefore, Cygnus\,X-1 is only detected while displaying persistent radio-jets, as claimed by \cite{Malyshev} and confirmed afterwards by \cite{Zdziarski2016}. Making use of the HS sample, we searched for orbital modulation (assuming ephemeris $T_{0}=52872.788$ HJD, \cite{Gies2008}). Orbital phases ($\phi$) were split into two bins, one centered at $\phi=0$, the superior conjunction of the compact object (0.75 $< \phi <$ 0.25) and other at the inferior conjunction (0.25 < $\phi$ < 0.75). Detection only occurred during superior conjunction (TS=31). Cygnus\,X-1 spectrum, from 60 MeV up to $\sim 20$ GeV, is well defined by a power law with photon index $\Gamma=2.3\pm0.1$ and normalization factor of $N_{0}=(5.8\pm0.9)\times 10^{-13}$ MeV$^{-1}$ cm$^{-2}$ s$^{-1}$, at an energy pivot of 1.3 GeV. Daily basis analysis was also performed, but no short-term flux variability was observed. The results between 0.1-20 GeV can be found in Figure \ref{CygX1LC}. 

\begin{center}
\begin{figure}	
\centering
		\includegraphics[width=0.7\linewidth]{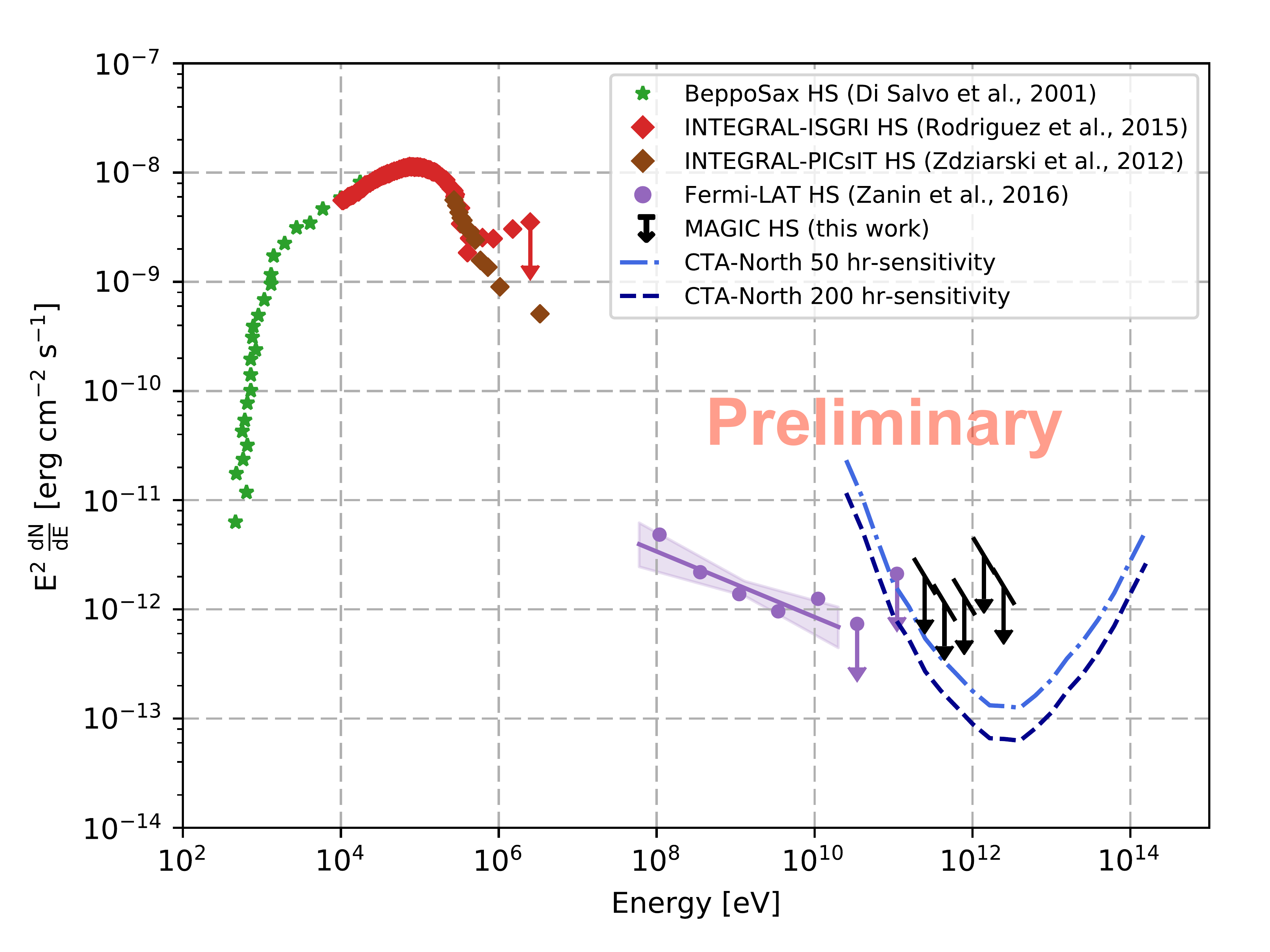}
		\caption{SED of Cygnus\,X-1. Soft X-rays from \protect\textit{BeppoSAX} are shown in green stars \protect\cite{DiSalvo2001}, while hard X-rays are taken from \textit{INTEGRAL}-ISGRI (red diamonds,\protect\cite{Rodriguez2015}) and \textit{INTEGRAL}-PICsIT (brown diamonds, \protect\cite{Zdziarski2012}). In the HE and VHE band, results presented in this proceeding obtained with \textit{Fermi}-LAT (violet points) and MAGIC (black ULs) are depicted. Sensitivity curves for CTA-North for 50 hours (https://www.cta-observatory.org/science/cta- performance/) and scaled to 200 hours of observations are shown in light blue and dark blue, respectively. No statistical errors are drawn, apart from the \textit{Fermi}-LAT butterfly.}
		\label{CygX1SED}
\end{figure}
\end{center}

With MAGIC, we searched for steady emission at energies above 200 GeV, making use of the total data set of $\sim 100$ hr. No significant excess was found, which led to an integral UL of $2.6\times 10^{-12}$ photons cm$^{-2}$ s$^{-1}$, assuming a power-law function with photon index $\Gamma=3.2$ (following former MAGIC results, \cite{Albert2007}). We also looked for gamma-ray emission at each X-ray state separately. In the HS, the source was observed for $\sim83$ hours between 2007-2011, which yielded no significant excess. Differential ULs are included in the spectral energy distribution (SED) shown in Figure \ref{CygX1SED}. Orbital phase-folded and daily analysis were also carried out, with no evidence of emission. Integral ULs in a night-by-night basis are depicted in Figure \ref{CygX1LC}. During SS, this microquasar was observed for $\sim 14$ hours in 2014. We searched for steady, orbital and short-term variability modulation, resulting in no detection. 

\subsection{Cygnus\,X-3}
\begin{center}
\begin{figure}
\centering
		\includegraphics[width=0.6\linewidth]{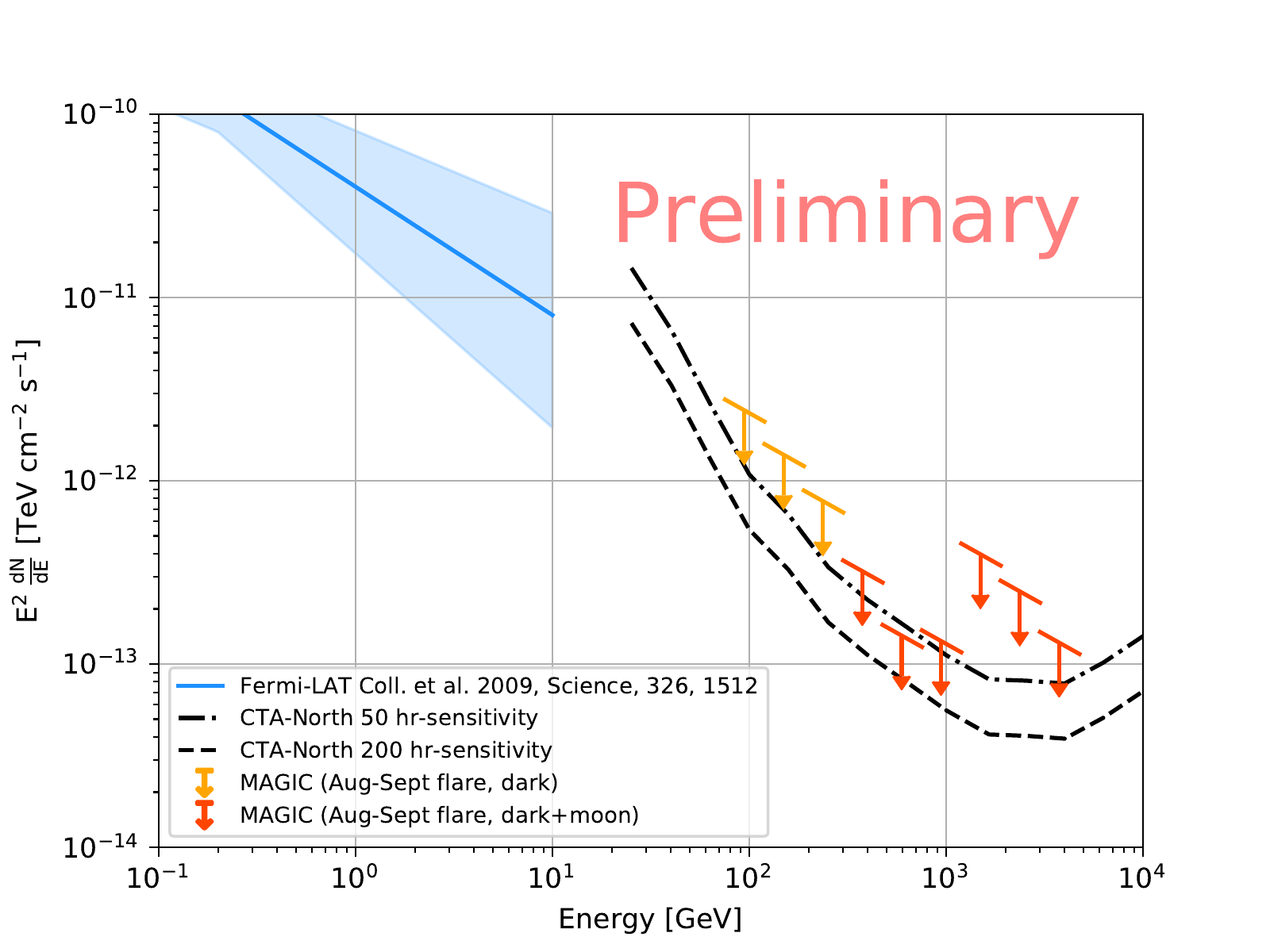}
		\caption{SED of Cygnus\,X-3. Blue butterfly corresponds to \textit{Fermi}-LAT spectrum during 2009 flare \protect\cite{Fermi2009}. MAGIC ULs for the August-September flare are represented in light orange ($\sim 52$ hours, dark data) and dark orange ($\sim 70$ hours, dark+moon data). Sensitivity curves for CTA-North for 50 hours (dot-dashed line) and 200 hours (dashed lines) observations are shown.}
		\label{CygX3SED}
\end{figure}
\end{center}

We searched for steady emission with the MAGIC telescopes, making use of the available $\sim 70$ hours. No excess was found at energies above 300 GeV (accounting for the energy threshold of the sample with the highest moonlight) nor 100 GeV (using $\sim 52$ hours of dark data, i.e. under absence of Moon). Differential ULs, assuming a power-law function with photon index $\Gamma=2.6$, are presented in Figure \ref{CygX3SED}. In this figure, \textit{Fermi}-LAT spectrum from \cite{Fermi2009} is taken, nevertheless \textit{Fermi}-LAT data for the August-September 2016 flare is currently being studied. No orbital (assuming ephemeris $T_{0}=2440949.892\pm 0.001$ JD, \cite{Singh2002}) or daily modulation was detected either.

\section{Discussion and conclusions}
HE and VHE gamma-ray emission were proposed in the literature from both leptonic and hadronic mechanisms (see e.g. \cite{BoschRamon2006,Romero2003}). Among these mechanism, the most efficient process seems to be a leptonic one, the IC. The target photons depend on the distance of the production site with respect to the compact object: close to it, thermal photons from the disk or synchrotron photons would dominate \cite{Romero2002,BoschRamon2006}; at a binary scales ($\sim R_{orb}$, the size of the system), IC would take place on stellar photons; and finally, gamma-ray emission could also be produced in the interaction between the jet and the medium (as seen in radio for Cygnus\,X-1, \cite{Gallo2005}). In the first two scenarios, gamma rays may suffer high absorption due to pair creation. 
\subsection{Cygnus\,X-1}
At the base of the jet, GeV photons would be absorbed by $\sim 1$ keV X-rays. Given the detection achieved with \textit{Fermi}-LAT, and following \cite{Aharonian} approach, we estimated the smallest region size for HE gamma-ray production at $2\times 10^{9}$ cm. The radius of the corona is $\sim 20-50~R_{g}\sim 5-10 \times 10^{7}$ cm \cite{Poutanen1998}, which allows us to conclude that the observed GeV emission is not originated in the corona, but most likely inside the jets. This scenario is reinforced by the fact that \textit{Fermi}-LAT detection only happens during HS. If the hint of orbital modulation here reported is finally confirmed, GeV emission must arrive from inside the jets and not from their interaction with the environment. Assuming so, we can set an UL on the largest distance of the production site at $< 10^{13}$ cm (few times $R_{orb}$ for this source). On the other hand, this flux variability is only expected if the radiative process that leads to GeV emission is anisotropic IC on stellar photons \cite{Khangulyan2014}. Given that the density of stellar photons is dominant over other photon fields at distances $>10^{11}$ cm, we place the GeV emitter at $10^{11}$--$10^{13}$ cm from the BH. 

On the other side, the MAGIC non-detection above 200 GeV allows us to discard jet-medium interaction as possible region for VHE emission above MAGIC sensitivity level, since these regions are not affected by photon-photon absorption. At binary scales this non-detection is less conclusive because of the pair production. Although VHE radiation is predicted in the models (see e.g. \cite{Pepe2015, Khangulyan2008}), several factors can prevent detection: low flux below MAGIC sensitivity even under negligible absorption \cite{Zdziarski2016}, no efficient acceleration on the jets or strong magnetic field. Nevertheless, transient events by relaxation of attenuation at some distance from the BH or extended pair cascade \cite{Zdziarski2009, BoschRamon2008} cannot be discarded. Transient emission related to discrete radio-emitting-blobs between HS and SS could also happen, as observed in the HE regime for Cygnus X-3. Hint of transient event was indeed reported previously by MAGIC \cite{Albert2007}. More sensitive instruments, like the future CTA (see Figure \ref{CygX1SED}), could provide interesting information on Cygnus\,X-1.

\subsection{Cygnus\,X-3}
Despite observing the source during strong radio and HE outbursts, no significant excess was found by MAGIC. One has to consider the extremely high absorption due to the WR, which may affect VHE gamma-ray emission. At energies above 300 GeV, the maximum absorption is produced by near-infrared (NIR) photons ($E_{target}\sim 1.7 $ eV). Following \cite{Aharonian2005}, absorption can be estimated as $\tau\sim \sigma_{\gamma \gamma}\cdot n_{NIR} \cdot R$, where $\sigma_{\gamma \gamma}\sim 1\times 10^{-25}$ cm$^{2}$ is the cross-section of the process, $n_{NIR}\sim L_{NIR}/(4 \pi R^{2} c E_{target})$ is the density of NIR photons and $R$ the size of the emitting region. Assuming the $L_{NIR}$ to be the bolometric luminosity,  $L_{NIR}=10^{38}$ erg s$^{-1}$, the absorption is not negligible until a radius $R\sim 10^{13}$ cm, i.e. outside the binary scale ($R_{orb,CygX3}\sim2.5\times 10^{11}$ cm). Given the MAGIC non-detection, acceleration up to VHE could still happen inside the jets at a distance $\lesssim 10^{13}$ cm, maybe related to the HE emission site (produced at $>10^{11}$ cm to avoid absorption by X-rays). On the other hand, MAGIC observed the source simultaneously with the strongest radio flare (at 9.5 Jy on MJD 57651), being the MAGIC significance for this day compatible with background. This could reinforce the idea that VHE gamma rays, if produced, are originated inside the binary scale and not at the radio-emitting regions of the jets far from the compact object. Note, however, that the amount of time observed during strong radio flares is very limited.

Figure \ref{CygX3SED} shows the Cygnus\,X-3 SED with the results at VHE during the 2016 flare, along with \textit{Fermi}-LAT spectrum taken from the 2009 flare \cite{Fermi2009}. As mentioned above, dedicated \textit{Fermi}-LAT analysis for the August-Sept 2016 flare is currently being performed. Our constraining ULs are also put in context with the CTA-North sensitivity curve for 50 hours of observations\footnote{Taken from (https://www.cta-observatory.org/science/cta- performance/} and the scaled one for 200 hours.


\begin{thebibliography}{99}
\bibitem{Orosz2011} Orosz J. A., et al., 2011, ApJ, 742, 84
\bibitem{Stirling2001} Stirling A. M., et al., 2001, MNRAS, 327, 1273
\bibitem{Gallo2005} Gallo E., et al., 2005, Nature, 436, 819
\bibitem{Brocksopp1999a} Brocksopp C., et al., 1999b, A\&A, 343, 861
\bibitem{Wen1999} Wen L., et al., 1999, ApJ, 525, 968
\bibitem{Brocksopp1999b} Brocksopp C., et al., 1999a, MNRAS, 309, 1063
\bibitem{Szostek2007} Szostek A., et al., 2007, MNRAS, 375, 793
\bibitem{Done2007} Done C., Gierli{\'n}ski M., Kubota A., 2007, A\&ARv, 15, 1
\bibitem{Malyshev} Malyshev D., et al., 2013, MNRAS, 434, 2380
\bibitem{AGILE2010Sabatini} Sabatini S., et al., 2010, ApJ, 712, L10
\bibitem{AGILE2010Bulgarelli} Bulgarelli A., et al., 2010, The Astronomer's Telegram, 2512
\bibitem{AGILE2013} Sabatini S., et al., 2013, ApJ, 766, 83
\bibitem{Albert2007} Albert J., et al., 2007, ApJ, 665, L51
\bibitem{Stark2003} Stark, M. J. et al. 2003, ApJ, 587, L101
\bibitem{Hanson2000} Hanson, M. M., et al. 2000, ApJ, 541, 308
\bibitem{Szostek2008} Szostek, A., et al. 2008, MNRAS, 388, 1001
\bibitem{Tavani2009} Tavani, M., et al. 2009, Nature, 462, 620
\bibitem{Fermi2009} Fermi LAT Collaboration, et al. 2009, Science, 326, 1512
\bibitem{PerformanceFermi} Atwood, W. B., et al., The Astrophysical Journal 697, 1071--1102 (2009)
\bibitem{Zanin2016} Zanin R., et al., 2016, A\&A, 596, A55
\bibitem{Aliu2009} Aliu E., et al., 2009, Astroparticle Physics, 30, 293
\bibitem{Alecksic2012} Aleksi\'c J., et al., 2012b, Astroparticle Physics, 35, 435
\bibitem{Alecksic2016}  Aleksi\'c J., et al., 2016b, Astroparticle Physics, 72, 76
\bibitem{FernandezBarral2017} MAGIC Collaboration et al., 2017, submitted to MNRAS
\bibitem{Zanin2013} Zanin R., et al., 2013, in Proc. of the 33st ICRC, Rio de Janeiro, Brasil.
\bibitem{AleksicLikelihood} Aleksi\'c J., et al., 2012a, J. Cosmology Astropart. Phys., 10, 032
\bibitem{RadioATel} Trushkin, S. A., et al. 2016a, The Astronomer's Telegram, 9416
\bibitem{FermiATel} Cheung, C. C. et al. 2016, The Astronomer's Telegram, 9502
\bibitem{MoonPerformance} MAGIC Collaboration, et al., 2017, ArXiv e-prints [1704.00906]
\bibitem{Rolke2005} Rolke, W. A., et al. 2005, Nuclear Instruments and Methods in Physics Research A, 551, 493
\bibitem{Gringberg2013} Grinberg V., et al., 2013, A\&A, 554, A88
\bibitem{Zdziarski2016} Zdziarski A. A., et al., 2016, ArXiv [1607.05059]
\bibitem{Gies2008} Gies D. R., et al., 2008, ApJ, 678, 1237
\bibitem{DiSalvo2001} DiSalvoT., et al., 2001, ApJ, 547, 1024
\bibitem{Rodriguez2015} Rodriguez J., et al., 2015, ApJ, 807, 17
\bibitem{Zdziarski2012} Zdziarski A. A., et al., 2012, MNRAS, 423, 663
\bibitem{Krimm2013} Krimm H. A., et al., 2013, ApJS, 209, 14
\bibitem{Matsuoka2009} Matsuoka M., et al., 2009, PASJ, 61, 999
\bibitem{Singh2002} Singh, N. S., et al. 2002, A\&A, 392, 161
\bibitem{BoschRamon2006} Bosch-Ramon V., et al., 2006, A\&A, 447, 263
\bibitem{Romero2003} Romero G. E., et al., 2003, A\&A, 410, L1
\bibitem{Romero2002} Romero G. E., et al., 2002, A\&A, 393, L61
\bibitem{Aharonian}  Aharonian, F. A., et al., apss 115, 201--225 October (1985)
\bibitem{Poutanen1998} Poutanen J. and Coppi P. S., Physica Scripta Volume T 77, p. 57 (1998)
\bibitem{Khangulyan2014} Khangulyan D., et al., The Astrophysical Journal 783, p. 100 (2014).
\bibitem{Pepe2015} Pepe C., Vila G. S., Romero G. E., 2015, A\&A, 584, A95
\bibitem{Khangulyan2008} Khangulyan D., et al., 2008, MNRAS, 383, 467
\bibitem{Zdziarski2009} Zdziarski A. A., et al., 2009, MNRAS, 394, L41
\bibitem{BoschRamon2008} Bosch-Ramon V., et al., 2008, A\&A, 489, L21
\bibitem{Aharonian2005} Aharonian, F., et al. 2005, Science, 309, 746

\end{thebibliography}
\end{document}